\documentclass[floatfix, preprint, showpacs, showkeys, preprintnumbers, nofootinbib, superscriptaddress]{revtex4-1}
\usepackage{comment}
\usepackage{amssymb}
\usepackage{color}
\usepackage{rotating}
\usepackage{amsmath}
\usepackage{ulem}
\usepackage{overpic}
\usepackage{graphicx}
\usepackage{dcolumn}
\usepackage{bm}
\usepackage{epstopdf}
\usepackage{chngpage}
\usepackage{multirow}
\usepackage{slashed}
\usepackage{indentfirst}
\usepackage{booktabs}
\usepackage{amssymb,amsfonts}
\usepackage{amsmath}
\usepackage{color}
\usepackage[colorlinks,citecolor=blue,anchorcolor=blue,linkcolor=blue,hypertex,breaklinks=true]{hyperref}
\usepackage{CJKutf8}

\begin{document}
\begin{CJK}{UTF8}{<font>}
\title{Local variations of charge radii for nuclei with even $Z$ from 84 to 120}
\author{Rong An}
\email[ ]{anrong@brc.ac.cn}
\affiliation{Key Laboratory of Beam Technology of Ministry of Education, Institute of Radiation Technology, Beijing Academy of Science and Technology, Beijing 100875, China}
\affiliation{Key Laboratory of Beam Technology of Ministry of Education, College of Nuclear Science and Technology, Beijing Normal University, Beijing 100875, China}
\affiliation{CAS Key Laboratory of High Precision Nuclear Spectroscopy, Institute of Modern Physics, Chinese Academy of Sciences, Lanzhou 730000, China}

\author{Xiao-Xu Dong}
\affiliation{School of Physics, Beihang University, Beijing 100191, China}

\author{Li-Gang Cao}
\email[ ]{caolg@bnu.edu.cn}
\affiliation{Key Laboratory of Beam Technology of Ministry of Education, College of Nuclear Science and Technology, Beijing Normal University, Beijing 100875, China}
\affiliation{Key Laboratory of Beam Technology of Ministry of Education, Institute of Radiation Technology, Beijing Academy of Science and Technology, Beijing 100875, China}

\author{Feng-Shou Zhang}
\email[Corresponding author: ]{fszhang@bnu.edu.cn}
\affiliation{Key Laboratory of Beam Technology of Ministry of Education, Institute of Radiation Technology, Beijing Academy of Science and Technology, Beijing 100875, China}
\affiliation{Key Laboratory of Beam Technology of Ministry of Education, College of Nuclear Science and Technology, Beijing Normal University, Beijing 100875, China}
\affiliation{Center of Theoretical Nuclear Physics, National Laboratory of Heavy Ion Accelerator of Lanzhou, Lanzhou 730000, China}


\begin{abstract}
 Pronounced changes of nuclear charge radii provide a stringent benchmark on the theoretical models and play a vital role in recognizing various nuclear phenomena.
  In this work, the systematic evolutions of nuclear charge radii along even $Z$=84-120 isotopic chains are firstly investigated by the recently developed new ansatz under the covariant density functional.
  The calculated results show that the shell closure effects of nuclear charge radii appear remarkably at the neutron numbers $N=126$ and 184. Interestingly, the arch-like shapes of charge radii between these two strong neutron closed shells are naturally observed. Across the $N=184$ shell closure, the abrupt increase in charge radii is still evidently emerged. In addition, the rapid raise of nuclear charge radii from the neutron numbers $N=138$ to $N=144$ is disclosed clearly in superheavy regions due to the enhanced shape deformation.
\end{abstract}


\maketitle
\section{INTRODUCTION}\label{sec0}
The precise determination of charge radii of nuclei far away from the $\beta$-stability line is a hot course in recent studies.
The knowledge of nuclear size plays an important role not only in searching for new physics beyond standard model (SM)~\cite{RevModPhys.90.025008} but also in serving as input quantities for astrophysical study~\cite{ARNOULD2020103766}.
Moreover, reliable descriptions of nuclear charge radii can reflect various nuclear structure phenomena and provide a stringent constraint on the equation of state (EoS) of isospin asymmetric nuclear matter~\cite{PhysRevLett.119.122502,PhysRevC.97.014314,PhysRevLett.127.182503,PhysRevResearch.2.022035}.

Fruitful experimental data show that the sizable local variations of nuclear charge radii are generally observed throughout the whole periodic table.
Remarkably in the calcium isotopes, nuclear charge radii feature a distinct aspect of odd-even staggering (OES) between the neutron magic numbers $N=20$ and $28$~\cite{ANGELI201369,LI2021101440}.
The same scenarios can also be found in the potassium isotopes, but the amplitudes are clearly weakening due to the last unpaired proton.
Toward proton-rich region, odd-even oscillation of charge radii still occurs in the calcium isotopes~\cite{Miller2019}.
Furthermore, the OES of charge radii is more pronounced in the neutron-deficient lead isotopes~\cite{ANSELMENT1986471}.
Also, this arresting trend in charge radii is evidently observed in the mercury~\cite{Marsh:2018wxs,PhysRevC.99.044306,PhysRevC.104.024328} and bismuth~\cite{PhysRevC.95.044324,PhysRevLett.127.192501} isotopes owing to the shape staggering.

The other conspicuous feature is the abrupt change in nuclear charge radii along an isotopic chain.
One kind of these rapid increases is presented naturally across the traditional magic numbers, this shell effect leads to the well-known kink phenomena in the trend of charge radii at the neutron numbers $N=28$, $50$, $82$ and $126$~\cite{Ruiz2016,Koszorus2020mgn,Gorges2019,GarciaRuiz:2019cog,Bhuyan_2021}.
As mentioned in~\cite{Goodacre2021}, the origin of this feature is universal and independent of local microscopic behavior.
This attributes to the rather small isospin dependence of the spin-orbit interactions in the quantum nature of atomic nucleus~\cite{PhysRevLett.74.3744}.
The other sudden increase is associated with the shape-phase transitions, such as the onset of deformation at the $N=60$ and $N=90$ regions where the rapid raise of nuclear charge radii is occurred ~\cite{LALAZISSIS199635,PhysRevLett.54.1991,PhysRevLett.117.172502}.

The evolution of nuclear charge radii shows the inverted parabolic-like shape between neutron numbers $N=20$ and $28$ in the potassium and calcium isotopic chains, but the trend for the former is visibly reduced~\cite{ANGELI201369,LI2021101440}.
Significantly, this arch-like structure has been observed generally between the two completely filled shells, such as in the copper ($Z=29$)~\cite{deGroote}, cadmium ($Z=48$)~\cite{PhysRevLett.121.102501} and tin ($Z=50$)~\cite{Gorges2019} isotopes, etc.
In recent years, plenty of information about the charge radii of nuclei far away from the $\beta$-stability line has been compiled. These available data provide a rigorous benchmark and challenge for the nuclear structure models~\cite{Reinhard:2021gym,PhysRevC.101.021301}.

Heavy or superheavy element synthesis and the location of the ``island of stability" are at the forefront in nuclear physics.
With the development of high intensity ion beam facilities and the advanced detectors, extensive efforts have been devoted to the synthesis of superheavy nuclei (SHN) and then the landscape has been extended to the northeast of the nuclide chart in laboratories~\cite{PhysRevLett.109.162501,Oganessian2015,Oganessian_2017}, namely up to the $Z=118$ element.
Actually, the information about the bulk properties of superheavy nuclei is hardly known in terrestrial experiment, especially for charge radii.
Although plenty of methods have been undertaken to extract the nuclear size, such as high-energy elastic electron scattering ($e^{-}$)~\cite{PhysRev.92.978,PhysRevC.21.1426}, muonic atom X-rays ($u^{-}$)~\cite{Engfer:1973df,FRICKE1995177,BAZZI2011199}, highly-sensitive Collinear Resonance Ionization Spectroscopy (CRIS)~\cite{Cocolios:2013wpa,vernon2020laser}, optical isotopes shift (OIS) and K$_{\alpha}$ X-rays isotopes shift (K$_{\alpha}$IS)~\cite{ANGELI2004185}, etc, the research on the charge radii of SHN is sort of beyond the ability of the available experimental tools due to the quite short lifetimes.
This can also essentially explicate the low production of cross sections of superheavy nuclei in experimental and theoretical studies~\cite{Niu2020,zhang2018}.
The fundamental properties of heavy or superheavy nuclei play an essential role not only in emerging out of the understanding of astrophysical process, but in producing new heavy elements~\cite{niufei2021,Xin2021}.
Therefore, the quite accurate description of charge radii in heavy or superheavy regions is necessary.

Plenty of models including empirical and microscopic approaches are contributed to describing the nuclear size.
The average trend of nuclear charge radii is usually scaled with $A^{1/3}$ or $Z^{1/3}$ law due to the consideration of nuclear saturation property~\cite{Zhang:2001nt}.
It is noted that the local variations cannot be properly captured and the rather large discrepancies still occur against experimental data.
Then the advanced versions have been proposed by introducing the shell effects and the OES behaviors~\cite{PhysRevC.88.011301,Sheng:2015poa}.
For the unstable nucleus, the $\alpha$-decay or cluster and proton emissions properties can also provide an alternative opportunity in exploring the nuclear size~\cite{PhysRevC.87.024310,PhysRevC.87.054323,PhysRevC.89.024318,QIAN2016134,Qian2018}.
Recently developed Bayesian neural networks (BNNs) have been devoted to evaluate the accurate predictions of nuclear charge radii, in which the theoretical error bars can also be given~\cite{Utama_2016,RenZZ2020,PhysRevC.102.054323,Dong:2021aqg,Dong:2022wkd}. The kernel ridge regression (KRR) method has also been used to learn nuclear charge radii~\cite{zhang2022}.
Toward superheavy regions, the extrapolative power for these methods is limited because of the absence of the database in the validation procedure.

Mean field theory makes a considerable success in describing the fundamental properties of finite nuclei~\cite{RevModPhys.75.121}.
The charge density distributions are calculated self-consistently in the framework of energy density functionals (EDFs), such as the relativistic~\cite{Geng:2003pk,PhysRevC.82.054319,XIA20181,PhysRevC.102.024314,Zhang:2021ize} and non-relativistic density functionals~\cite{PhysRevC.82.035804,PhysRevLett.102.242501}.
As mentioned in~\cite{PhysRevLett.128.022502}, the local variations in charge radii, such as the shell effects and OES behaviors, cannot be reproduced well by Skyrme EDFs.
The non-relativistic Fayans density functional can overcome these deficiencies with considering the surface pairing interactions~\cite{Reinhard2017}.
Nevertheless, recent study~\cite{Goodacre2021} suggests that the pairing does not need to play a crucial role in describing both the shell closure at $N=126$ and the OES of charge radii in its vicinity for the mercury isotopes.
$Ab$ $initio$ many-body calculations with chiral effective field theory (EFT) interactions have also been employed to investigate the nuclear charge radii~\cite{BINDER2014119,PhysRevC.91.051301,PhysRevC.96.014303,PhysRevC.101.014318,Stroberg}. However, there are still some difficulties in heavy or superheavy regions because of the limited computational power.

As mentioned in~\cite{GarciaRuiz:2019cog}, nuclear charge radii present global and simple patterns along isotopic chains.
A recently developed ansatz~\cite{An:2020qgp} which phenomenologically takes into account the neutron-proton correlations correction can describe the shell closure effects and OES behaviors in charge radii. A systematic global investigation of nuclear charge radii in superheavy regions has not been performed by this modified method.
Hence the objective of this article is worth extracting the local variations in the root mean square (rms) charge radii of even $Z=84$-$120$ isotopes. The systematic trend of nuclear charge radii across the $N=126$ and 184 shell closures and the shape-phase transition regions should be paid more attention as well.

This paper is organized as follows. In section~2, we briefly describe the theoretical approach for describing the nuclear charge radii.
Section~3 is devoted to the results and discussions. A short summary and outlook are provided in section~4.
\section{Theoretical framework\label{Sec3}}
Many nuclear phenomena can be described successfully by the covariant density functional theory (CDFT)~\cite{Vretenar:2005zz,Zhao:2014lca,Liang:2014dma,zhang2007,Zhao:2012ck,PhysRevC.67.034312,PhysRevC.68.034323,PhysRevC.69.054303,
PhysRevC.82.011301,Meng:2005jv,Cao:2003yn,zhang2012,An:2020wcv,Zss22,KunWang15303}.
A recently developed ansatz in charge radii~\cite{An:2020qgp} has been built under the framework of CDFT, in which the model Lagrangian density is recalled as
\begin{eqnarray}
\mathcal{L}&=&\bar{\psi}[\mathrm{i}\gamma^\mu\partial_\mu-M-g_\sigma\sigma
-\gamma^\mu(g_\omega\omega_\mu+g_\rho\vec
{\tau}\cdotp\vec{\rho}_{\mu}+e\mathbf{A}_\mu)]\psi\nonumber\\
&&+\frac{1}{2}\partial^\mu\sigma\partial_\mu\sigma-\frac{1}{2}m_\sigma^2\sigma^2
-\frac{1}{3}g_{2}\sigma^{3}-\frac{1}{4}g_{3}\sigma^{4}\nonumber\\
&&-\frac{1}{4}\mathbf{\Omega}^{\mu\nu}\mathbf{\Omega}_{\mu\nu}+\frac{1}{2}m_{\omega}^2\omega_\mu\omega^\mu
+\frac{1}{4}c_{3}(\omega^{\mu}\omega_{\mu})^{2}-\frac{1}{4}\vec{R}_{\mu\nu}\cdotp\vec{R}^{\mu\nu}\nonumber\\
&&+\frac{1}{2}m_\rho^2\vec{\rho}^\mu\cdotp\vec{\rho}_\mu
+\frac{1}{4}d_{3}(\vec{\rho}^{\mu}\vec{\rho}_{\mu})^{2}-\frac{1}{4}\mathbf{F}^{\mu\nu}\mathbf{F}_{\mu\nu}.
\end{eqnarray}
Here, $M$, $m_{\sigma}$, $m_{\omega}$ and $m_{\rho}$ are the nucleon, $\mathbf{\sigma}$-, $\mathbf{\omega}$- and $\mathbf{\rho}$-meson masses, respectively.
In addition, the parameter sets $g_{\sigma}$, $g_{\omega}$, $g_{\rho}$, $g_{2}$, $g_{3}$, $c_{3}$, $d_{3}$ and $e^{2}/4\pi=1/137$ are the coupling constants for $\mathbf{\sigma}$, $\mathbf{\omega}$, $\mathbf{\rho}$ mesons, and photon, respectively. The field tensors for the vector mesons are $\mathbf{\Omega}_{\mu\nu}=\partial_{\mu}\omega_{\nu}-\partial_{\nu}\omega_{\mu}$, $\vec{R}_{\mu\nu}=\partial_{\mu}\vec{\rho}_{\nu}-\partial_{\nu}\vec{\rho}_{\mu}-g_{\rho}(\vec{\rho}_{\mu}\times\vec{\rho}_{\nu})$ and for the electromagnetic field $\mathbf{F}_{\mu\nu}=\partial_{\mu}\mathbf{A}_{\nu}-\partial_{\nu}\mathbf{A}_{\mu}$.

By combining the Dirac spinors $\psi$ and the variational principle, the stationary Dirac equation can be derived from the Lagrangian density.
It can be rewritten as follows,
\begin{eqnarray}
\hat{\mathcal{H}}\psi_{i}(\mathbf{r})=\epsilon_{i}\psi_{i}(\mathbf{r}).
\end{eqnarray}
These corresponding Dirac equations for nucleons are solved self-consistently by the expansion method with the axially symmetric harmonic oscillator basis~\cite{Geng:2003pk,Ring:1997tc}.
We perform calculations with modified linear constraint method in solving the Dirac equations~\cite{PhysRevC.89.014323},
\begin{eqnarray}
(\hat{\mathcal{H}}-\lambda\langle{\mathbf{Q}}\rangle)\psi_{i}(\mathbf{r})=\epsilon_{i}\psi_{i}(\mathbf{r}),
\end{eqnarray}
where $\lambda$ is the spring constant and $\mathbf{Q}$ is the intrinsic quadrupole moment.
The ground state properties of finite nuclei are obtained by constraining the quadrupole deformation parameter $\beta_{20}$. Those values of $\beta_{20}$ are changed from $-0.50$ to $0.50$ with the interval of $0.01$.

In order to reproduce the anomalous behaviors of nuclear charge radii along calcium isotopic chain, the semi-microscopic correction term has been introduced into the rms charge radius formula.
The modified part in this expression is firstly mandatory for describing the odd-even oscillations and the inverted parabolic-like shapes of nuclear charge radii between neutron numbers $N=20$ and $28$~\cite{An:2020qgp}. Meanwhile, it can also be used to describe the charge radii of the odd-$Z$ potassium isotopes~\cite{An:2021rlw}.
The modified mean-square charge radius has the following expression,
\begin{eqnarray}\label{coop1}
R_{\mathrm{ch}}^{2}=\frac{\int{r}^{2}\rho_{\mathrm{p}}(\mathbf{r})\mathrm{d}^{3}r}{\int\rho_{\mathrm{p}}(\mathbf{r})\mathrm{d}^{3}r}+0.64~\mathrm{fm^{2}}+\frac{a_{0}}{\sqrt{A}}|\Delta{\mathcal{D}}|~\mathrm{fm^{2}}.
\end{eqnarray}
Where the first term represents the charge density distributions of point-like protons and then the value of 0.64 fm$^2$ accounts for the finite size effects of proton~\cite{Gambhir:1989mp,Ring:1997tc}. For the last term, the quantity $a_{0}=0.22$ is an normalization parameter and $A$ represents the mass number of a nucleus. The expression $|\Delta{\mathcal{D}}|$ originates from the differences of Cooper pairs components between neutrons and protons~\cite{An:2020qgp}. For ease of discussion, the results obtained via equation~(\ref{coop1}) with and without considering the modified term $\Delta\mathcal{D}$ are denoted as RMF(BCS)* and RMF(BCS), respectively.

Moreover, the abrupt changes of nuclear charge radii across shell closures can also be reproduced well~\cite{An:2021seo}. The value of $\Delta{\mathcal{D}}$ is obtained by solving the state-dependent BCS equations with a zero-range $\delta$-force interactions~\cite{Geng:2003pk}. For odd-$A$ nuclei, the last single particle level occupied by the odd nucleon is blocked. The pairing strength $V_{0}=322.8$ MeV fm$^{3}$ is employed for all calculations~\cite{An:2020qgp}. The effective force NL3 is especially favorable to the calculation of charge radii along a long isotopic chain~\cite{PhysRevC.55.540}. Actually, parametrization set PK1 and NL3$^{*}$ cannot change the conclusions in quantitative level~\cite{An:2021seo}.

The evolution of charge radii is reproduced remarkably well including kinks at neutron magic numbers and OES behaviors. This modified item is explained phenomenologically by the neutron-proton correlations around Fermi surface. As demonstrated in~\cite{Miller:2018mfb}, neutron-proton interactions can influence the calculated rms charge radius, but it is invalid for some cases. Especially in our recent work~\cite{An:2021gxz}, this ansatz has almost no influence in describing some nuclei with open proton shell due to the strong coupling between different levels around Fermi surface, this is in accord with Miller's~\cite{Miller:2018mfb}.

As argued in~\cite{Perera:2021ztx}, the modified formula can improve the description of nuclear charge radii well, but it cannot be derived self-consistently from the conventional density functional theories (DFTs). Although the origin of the underlying physical meaning of this ansatz need to be further discussed, its predictive power is non-negligible and the influence on the neutron-proton ($np$) interactions should be paid more attention in describing charge radii. In addition, the Casten factor which is used to cover the shell effect in charge radii has the form $N_{p}N_{n}/(N_{p}+N_{n})$~\cite{PhysRevLett.58.658,Angeli_1991,PhysRevC.88.011301}, where the valence neutron ($N_{n}$)-proton ($N_{p}$) interactions have been contributed to the variations of nuclear size. This suggests that the neutron-proton interactions play an essential role in describing the fine structure of nuclear charge radii.

\section{Results and discussions\label{Sec4}}
\subsection{Odd-even staggering in nuclear charge radii}\label{third0}
As is well known, odd-even staggering (OES) behaviors in nuclear charge radii are generally observed throughout the nuclear chart~\cite{ANGELI201369,LI2021101440}. The source of OES behavior is demonstrated by various possible mechanisms~\cite{Perera:2021ztx}, such as the staggering oscillation in the occupation of the single-particle levels between odd and even isotopes~\cite{Goodacre2021} and the pairing contributions~\cite{Gorges2019}. In addition, OES in charge radii might be related to the evolution of nuclear deformation~\cite{SAKAKIHARA2001649}, etc. Although this physical source is complex, introducing OES term is indispensable in validation procedure~\cite{Dong:2021aqg}.

\begin{figure}[htbp]
\centering
\includegraphics[scale=0.4]{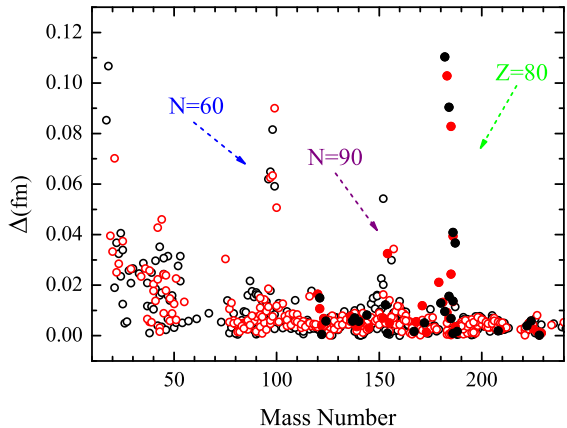}
   \caption{(Color online) The absolute values of odd-even staggering in nuclear charge radii (defined as $\Delta_{r}(N,Z)$) are obtained by the three-point formula~(\ref{eq3}), in which black open circles (solid circles) represent the positive values and the opposite sign is labeled by red open circles (solid circles). The solid circles represent the so-called abnormal odd-even staggering in charge radii. The deformation phase transition ($N=60$ and $90$) and shape staggering regions ($Z=80$) are highlighted by colored arrows. The available experimental data are taken from~\cite{ANGELI201369,LI2021101440}.} \label{fig0}
\end{figure}
As shown in~\cite{Reinhard2017,Borzov:2020lwt}, the three-point OES formula is employed to visually measure those local variations of nuclear charge radii which reads,
{\small \begin{eqnarray}\label{eq3}
\Delta_{r}(N,Z)=\frac{1}{2}[R(N-1,Z)-2R(N,Z)+R(N+1,Z)],
\end{eqnarray}}
where $R(N,Z)$ is the rms charge radius of a nucleus with neutron number $N$ and proton number $Z$.

In figure~\ref{fig0}, the absolute values for $\Delta_{r}(N,Z)$ are shown as a function of mass number $A$. These radius differences $\Delta_{r}(N,Z)$ are extracted from the available experimental data of charge radii~\cite{ANGELI201369,LI2021101440}. The amplitudes of OES behaviors are gradually weaken with the increasing mass number.
However, around $N=60$ and $90$ regions, the enlarged amplitudes are indicated due to the sudden onset of strong static deformations~\cite{LALAZISSIS199635,PhysRevLett.54.1991,PhysRevLett.117.172502}. This strong peak can also be clearly presented in the Brix-Kopfermann diagram, which provides a widely applicable prototype for recognizing the onset of deformation throughout heavy nuclei~\cite{RevModPhys.30.517}. Moreover the large amplitudes of OES appear at the neutron-deficient Hg ($Z=80$) isotopes due to the shape staggering~\cite{Marsh:2018wxs,PhysRevC.104.024328}, namely a large and abrupt shape transition from prolate to oblate and again back to prolate. This phenomenon is attributed to the descent of the $h_{9/2}$ ``intruder" configuration which destroys the $Z = 82$ gap, and therefore increases the effective number of valence protons~\cite{GOODMAN19771}. The latest observation suggests that the noticeably OES can also be encountered in neutron-deficient bismuth isotopes~\cite{PhysRevLett.127.192501}.

The empirical averaging energy gap on mass number obtained by three-point formula shows the similar trends but with larger magnitudes, and roughly equals to $12/\sqrt{A}$~\cite{P.Ring}. The modified term in rms charge radius formula includes similar expression $\propto{1/\sqrt{A}}$~\cite{An:2020qgp}, which has been used to capture the OES behaviors in nuclear charge radii. It is mentioned that the pairing energy gap decreases with mass number $A$ approximately as $\propto{1/\sqrt{A}}$~\cite{JANECKE200323}. Notablyㄛthere is a common suppression factor ${1/\sqrt{A}}$ in the aforementioned models.

In general, normal OES effects disclose that the charge radii of odd-$N$ isotopes are smaller than those of their even-$N$ neighbors.
This can be characterized through the values of $\Delta_{r}(N,Z)$ shown as open circles in figure~\ref{fig0}. The negative sign corresponds to even-$N$ isotopes and the positive sign represents the odd-$N$ case. On the contrary, the so-called abnormal OES effects of nuclear charge radii show an opposite trend along isotopic chain as drawn in figure~\ref{fig0}, which are labeled by solid circles~\cite{ANGELI201369}.
This inverse OES in charge radii, which may be associated with the presence of octupole collectivity, is observed apparently in astatine isotopes (See~\cite{Barzakh2019} and references therein). As mentioned in~\cite{Lievens:1995iz}, the inverted OES is interpreted as an effect of polarization. The unpaired neutron polarizes the soft even-even core towards the pronounced octupole deformation~\cite{AHMAD1988244}. Similar arguments have been used to interpret the anomalous OES in the europium isotopes with $N = 89$-$92$~\cite{Alkhazov1990}. As argued in~\cite{SAKAKIHARA2001649}, pair mean field as well as the gap potential are both responsible for the occurrence of OES in nuclear size. Since it is still a long-standing and open topic in nuclear physics.

\subsection{Charge radii of the even-$Z$=84-96 isotopes}\label{third1}
\begin{figure}[htbp]
\centering
\includegraphics[scale=0.45]{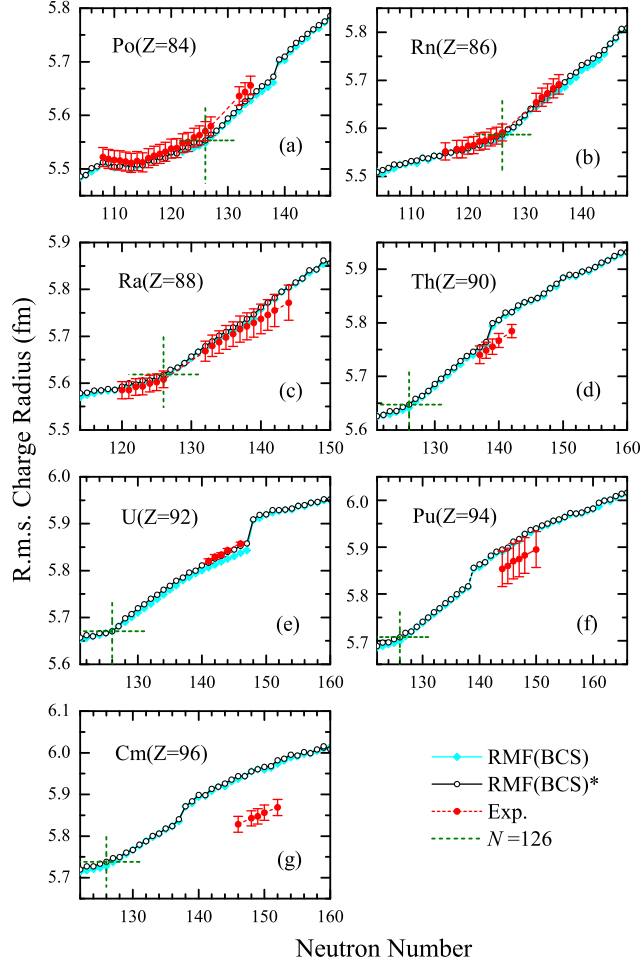}
   \caption{Charge radii of polonium (a), radon (b), radium (c), thorium (d), uranium (e), plutonium (g) and curium (h) isotopes are obtained by the RMF(BCS) (solid diamond) and the modified RMF(BCS)* (open circle) approaches, respectively. The experimental data are taken from~\cite{ANGELI201369}.} \label{fig1}
\end{figure}

To facilitate the quantitative comparison of the experimental results with these theoretical calculations, the root-mean-square (rms) charge radii of the polonium ($Z=84$), radon ($Z=86$), radium ($Z=88$), thorium ($Z=90$), uranium ($Z=92$), plutonium ($Z=94$) and curium ($Z=96$) isotopes are shown in figure~\ref{fig1}.
One can find that the calculated rms charge radii compare well against the experimental data for polonium, radon, radium, thorium and uranium isotopes. For plutonium isotopes, the results obtained by the new ansatz are overlapping partially with the current uncertainty in~\cite{ANGELI201369}. However, the calculated results deviate significantly from the experimental data along curium isotopic chains.
 Our calculations are consistent with results obtained by the deformed relativistic Hartree-Bogoliubov theory in continuum (DRHBc) with the density functional PC-PK1~\cite{Zhang2022ADNDT}.
It can be found that the results obtained by the RMF(BCS) and RMF(BCS)* approaches are comparable except for $^{232-239}$U isotopes. As suggested in~\cite{An:2021gxz}, these largely differences come from the modified term $\Delta{D}$.

As shown in figure~\ref{fig1}, the slopes of the rms charge radii as a function of neutron number are similar in the neutron range $N=126-138$. The rapid increases in rms charge radii can be found remarkably across the neutron number $N=138$ along polonium, thorium, plutonium and curium isotopic chains. Actually, this scenario can also be found in radon and radium isotopes but with small amplitudes. As mentioned in~\cite{Perera:2021ztx}, the different sequences of the occupation of the single-particle levels lead to the changes of charge radii. In which the subshell $1i_{11/2}$ is completely filled ($N=138$) and then the added neutron will be filled in subshell $2g_{9/2}$ up to $N=148$. In our calculations, the calculated quadrupole deformations along even $Z=84-96$ isotopic chains are almost spherical from $N=126$ to $N=138$.
Across $N=138$ subshell, the rapid raise in rms charge radii attributes to the increasing deformation parameter, namely $\beta_{20}>0.20$. In uranium isotopes, the jump in charge radii is disappeared due to the calculated values of $\beta_{20}$ are almost close to zero from $N=121-147$. Across the $N=148$, the rapid change of the rms charge radius can also be observed evidently in uranium isotopes, in which the deformation parameter $\beta_{20}$ equals to $0.30$.

The shell effect in charge radii is still emerged naturally across the $N=126$ shell closure.
The calculated results suggest that quadrupole deformation has an influence on the systematic global evolution of nuclear charge radii.
This leads to the complicated interplay of the occupations of different single-particle states around Fermi surface. Particularly, it can be seen from the uranium isotopes where the abrupt raise in rms charge radii is vanished across the neutron number $N=138$ but occurs at $N=148$. In addition, the sequence of the occupation of the single-particle subshells is influenced by the chosen parameter set~\cite{Perera:2021ztx}.


\subsection{Charge radii of the even $Z=98$-$120$ isotopes}\label{third2}
\begin{figure*}[htbp!]
\centering
\includegraphics[scale=0.80]{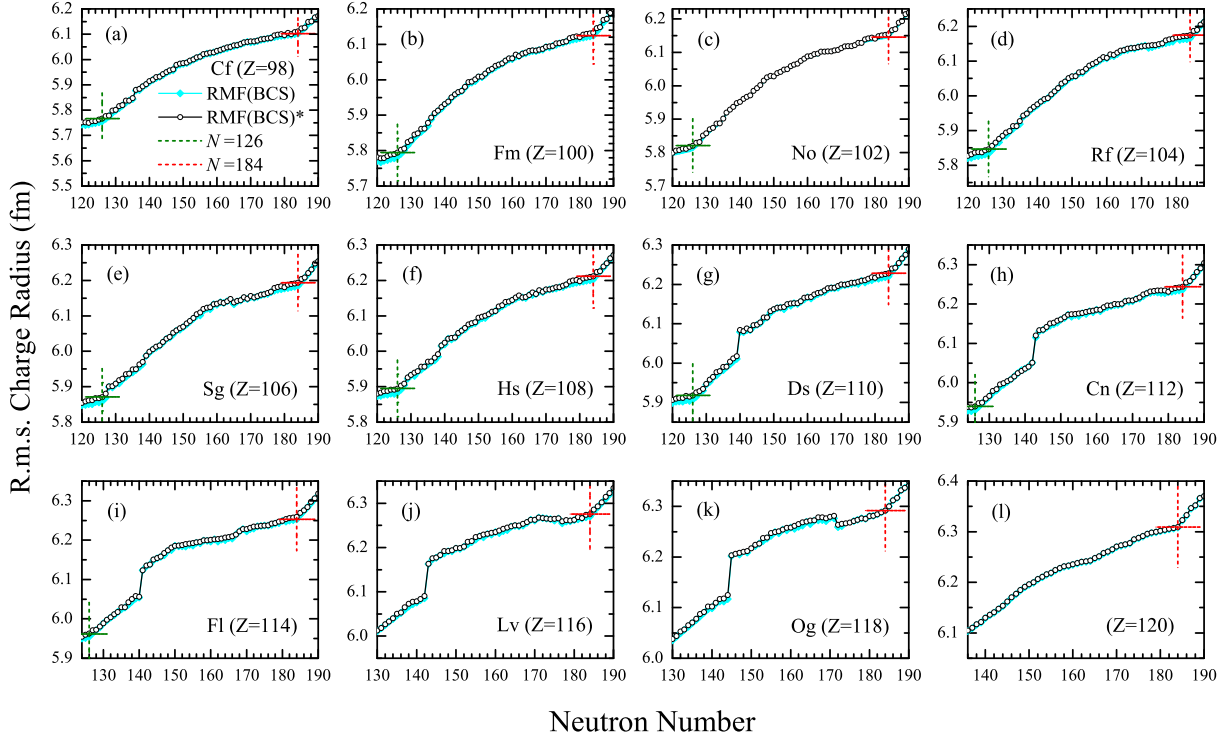}
   \caption{Charge radii of californium (a), fermium (b), nobelium (c), rutherfordium (d), seaborgium (e), hassium (f), darmstadtium (g), copernicium (h), flerovium (i), livermorium (j), oganesson (k) and $Z=120$ (l) isotopes are drawn by the RMF(BCS) (solid diamond) and the modified RMF(BCS)* (open circle) approaches, respectively.} \label{fig2}
\end{figure*}
As shown in figure~\ref{fig1}, this ansatz can provides a reliable descriptions of nuclei charge radii for even $Z=84$-$96$ isotopes.
In order to further evaluate the global evolution of charge radii in superheavy regions, especially the abrupt raise due to the shape-phase transition and shell effect, nuclear sizes along even $Z=98$-$120$ isotopic chains are calculated by this ansatz.
Although the amplitudes of OES decrease dramatically with the increasing mass number as shown in figure~\ref{fig0}, the apparently staggering behaviors are expected to occur in visually quantitative level.

In figure~\ref{fig2}, nuclear rms charge radii of even $Z=98$-$120$ isotopes are calculated by this ansatz.
The results obtained by the RMF(BCS) method are also shown for comparison. In our calculations, the values performed by RMF(BCS)* method deviate slightly from the RMF(BCS) approach. The local variations of charge radii along even $Z=98-120$ isotopic chains are similar for these two approaches.
As discussed before, charge radii vary smoothly toward neutron closed-shells, and then a larger increase occurs through the filling of the new open shells~\cite{Ruiz2016,Gorges2019,GarciaRuiz:2019cog}.
For even $Z=98$-$114$ isotopic chains, the remarkably abrupt changes in charge radii are observed clearly across neutron shell closure $N=126$.
As shown in~\cite{KIRSON200829,ZHANG2005106,Oganessian2015}, neutron number $N=184$ is suggested to be the magic number.
Therefore, as shown in figure~\ref{fig2}, results obtained by RMF(BCS)$^{*}$ approach present a rapidly increasing phenomenon across the main neutron shell of $N = 184$ along $Z=98$-$120$ isotopic chains.

The RMF(BCS)$^{*}$ model can reproduce the shell effect in nuclear size fairly well for K, Ca, Cu and In isotopes~\cite{An:2020qgp,An:2021rlw,An:2021seo}, etc, especially the inverted parabolic-like behaviors between the two closed-shells. This is an indicator of the shell effect along a long isotopic chain. Between these two $N=126$ and $184$ filled-shells, the inverted parabolic-like behaviors of charge radii can be seen evidently but with lower amplitudes, which is shown in figures~\ref{fig2}~(a)-(f).

Other interesting phenomena that nuclear charge radii appear to increase with surprisingly leaping slopes, which are observed remarkably around some specific neutron numbers. As suggested in~\cite{PhysRevLett.117.172502}, the abrupt changes of nuclear charge radii correspond to the onset of deformation as well. Besides the abrupt changes at neutron shell closure, the changes in nuclear deformation are adequate to cause the observed kink phenomena~\cite{SAKAKIHARA2001649}. For the Cf $(Z=98)$ isotopes, the slightly abrupt increase in charge radii can be found at $N=136$. However, along Fm ($Z=90$), Nd ($N=102$), Rf $(Z=104)$, Sg ($Z=106$) and Hs ($Z=108$) isotopic chains, the sharp increases cannot be found in nuclear rms charge radii. At the neutron number $N=140$, the abrupt increase in nuclear charge radii can also be predicted along Ds ($Z=110$) and Fl $(Z=114)$ isotopic chains. For Lv ($Z=116$) isotopes, this phenomenon occurs at neutron number $N=142$. However, this sharp changes are encountered at the neutron number $N=144$ along Cn $(N=112)$ and Og $(N=118)$ isotopes. It is found that the rms charge radii in the Og isotopes are decreased slightly across the neutron number $N=172$. This stems from the shape-phase transition where the quadrupole deformation $\beta_{20}$ closes to zero. This is in accord with the results calculated by PC-PK1 density functional~\cite{Zhang2022ADNDT}.

As discussed above, the nuclear shape is abruptly increased from spherical to deformed around $N=140$, 142 and $144$ regions. The corresponding quadrupole deformation parameters are $\beta_{20}\approx0.30$, and these are consistent with~\cite{MOLLER20161}. This means deformation effect plays a crucial role in defining the properties such as nuclear sizes and isotope shifts.

\section{Summary}\label{Sec5}
In this work, the recently developed ansatz is firstly employed to investigate the systematic evolution of nuclear charge radii along even-$Z=84$-$120$ isotopic chains. The abrupt changes of nuclear charge radii at $N=126$ closure shell are produced apparently by RMF(BCS)$^{*}$ approach due to the rather small isospin dependence of the spin-orbit interactions. In our calculations, the remarkably abrupt changes in charge radii are also observed naturally across the neutron-closure shell $N=184$. Besides, the inverted parabolic-like behaviors between these two strong neutron numbers $N=126$ and $N=184$ are also shown but with slight amplitudes.

Actually, nuclear charge radii are influenced by various mechanisms, such as high order moment~\cite{PhysRevC.101.021301,PhysRevC.104.024316}, spherical-deformed phase transition~\cite{Reinhard:2021gym}, pairing components~\cite{Reinhard2017}, ${N}_{\mathrm{p}}{N}_{\mathrm{n}}$ scheme that describes the proton-neutron interactions~\cite{PhysRevC.33.1819}, etc. As shown in~\cite{FAYANS200049} that the density dependent pairing may also induce sizeable staggering and kinks in the evolution of the nuclear charge radii. As shown in figure~\ref{fig0}, the abrupt changes of nuclear charge radii are indicated around $A=100$, $150$ and $190$ regions. Results obtained by RMF(BCS)$^{*}$ approach show a remarkable similarity that deformation phase transition regions are also shown around $N=138$ to $144$ regions. These indicators for respective transition regions can be used to guide further test.

\section{Acknowledgements}
The project is funded by the Key Laboratory of High Precision Nuclear Spectroscopy, Institute of Modern Physics, Chinese Academy of Sciences. This work is also supported partly by the National Natural Science Foundation of China under Grants No. 12135004, No. 11635003, No. 11961141004 and No. 12047513. L.-G. C. acknowledges the support of the National Natural Science Foundation of China under Grants No. 11975096 and the Fundamental Research Funds for the Central Universities (2020NTST06).

\vspace*{2mm}

\begin{small}
\end{small}
\end{CJK}
\end{document}